\documentclass[aps,pra,11pt,longbibliography]{revtex4-1}
\usepackage[latin1]{inputenc} 
\usepackage[T1]{fontenc}

\usepackage{graphicx}
\usepackage{framed}
\usepackage{esvect}
\usepackage{amsmath,amssymb,color,bm}
\usepackage{mathrsfs}
\usepackage[dvipsnames]{xcolor}

\newcommand{\rev}[1]{{\color{black} {#1}}}

\newcommand\tb[1]{\textbf{#1}}

\newcommand\tn[1]{\textnormal{#1}}

\newcommand\mc[1]{\mathcal{#1}}

\newcommand\beq{\begin{equation}}
\newcommand\eeq{\end{equation}}
\newcommand\beqa{\begin{eqnarray}}
\newcommand\eeqa{\end{eqnarray}}

\newcommand\e[1]{\cdot 10^{#1}}

\def\lambe{\lambda_{\rm e}}

\def\lambf{\lambda_{\rm h}}

\def\lambef{\lambda_{\rm he}}
\def\Lambef{\Lambda_{\rm he}}

\def\taue{\tau_{\rm e}}

\def\Pe{\mc{P}_{\rm e}}
\def\Pf{\mc{P}_{\rm h}}

\def\L{\mc{L}}
\def\Cef{\mc{C}_{\rm he}}

\def\m{_{\rm max}}

\def\.{\cdot}

\def\1{^{-1}}
\def\2{^{-2}}
\def\3{^{-3}}

\begin{document}
\title{Hydroelectric energy conversion of waste flows through hydroelectronic drag}

\author{Baptiste Coquinot${}^{1}$}
\author{Lyd\'eric Bocquet${}^1$}\email{lyderic.bocquet@ens.fr}
\author{Nikita Kavokine${}^{2,3,4}$}\email{nikita.kavokine@epfl.ch}

\affiliation{${}^1$\!\!\! Laboratoire de Physique de l'\'Ecole Normale Sup\'erieure, ENS, Universit\'e PSL, CNRS, Sorbonne Universit\'e, Universit\'e Paris Cit\'e, 24 rue Lhomond, 75005 Paris, France\\
${}^2$\!\!\! Max Planck Institute for Polymer Research, Ackermannweg 10, 55128 Mainz, Germany \\
${}^3$ \!\!\!  Center for Computational Quantum Physics, Flatiron Institute, 162 5$^{th}$ Avenue, New York, NY 10010, USA \\
${}^4$ \!\!\!  The Quantum Plumbing Lab (LNQ), \'Ecole Polytechnique F\'ed\'erale de Lausanne (EPFL), Station 6, CH-1015 Lausanne, Switzerland}

\date{\today}

\begin{abstract}
\textbf{Hydraulic energy is a key component of the global energy mix, yet there exists no practical way of harvesting it at small scales, from flows with low Reynolds number. This has triggered a search for alternative hydroelectric conversion methodologies, leading to unconventional proposals based on droplet triboelectricity, water evaporation, osmotic energy or flow-induced ionic Coulomb drag. Yet, these approaches systematically rely on ions as intermediate charge carriers, limiting the achievable power density. Here, we predict that the kinetic energy of small-scale "waste" flows can be directly and efficiently converted into electricity thanks to the hydro-electronic drag effect, by which an ion-free liquid induces an electronic current in the solid wall along which it flows. This effect originates in the fluctuation-induced coupling between fluid motion and electron transport.
We develop a non-equilibrium thermodynamic formalism to assess the efficiency of such hydroelectric energy conversion, dubbed hydronic energy. We find that hydronic energy conversion is analogous to thermoelectricity, with the efficiency being controlled by a dimensionless figure of merit. However, in contrast to its thermoelectric analogue, this figure of merit combines independently tunable parameters of the solid and the liquid, and can thus significantly exceed unity. Our findings suggest new strategies for blue energy harvesting without electrochemistry, and for waste flow mitigation in membrane-based filtration processes.}
\end{abstract}
\maketitle

\textbf{Significance statement}. 
\emph{At large scales, where fluid dynamics are governed by inertia, a turbine can convert this inertia into electricity with an efficiency in excess of 90\%. At the smaller scales where viscous effects dominate, turbines are no longer efficient, and no satisfactory alternative has been identified so far.  Yet, significant amounts of energy are lost to viscous flows in membrane-based filtration and osmotic energy conversion. Here, we propose a new physical principle for nanoscale hydroelectric energy conversion -- dubbed "hydronic energy" -- which exploits the recently discovered fluctuation-induced quantum friction phenomenon. We develop a theoretical framework to assess its efficiency, and demonstrate its pertinence for generating electricity from liquid flows in the viscous regime. }

Since its invention by Hero of Alexandria two thousand years ago, the turbine has been the tool of choice for converting the kinetic energy of a fluid flow into useful work. It is now a mature technology, with the hydroelectric conversion efficiency of modern turbines reaching over 90\% \cite{Gordon2001}. However, turbines only function at macroscopic  scales and with sufficiently fast flows. When the Reynolds number decreases below the turbulence transition, the turbine efficiency plummets \cite{Deam2008,Lemma2008}.
Thus, from an energetic standpoint, low Reynolds number flows are "waste flows", in the same way as temperatures below 100$^{\circ}$C are low-grade waste heat, still out of reach in terms of industrial energy recovery.
Waste flows systematically arise in membrane-based filtration processes, as they involve liquids flowing through nanoscale pores \cite{McGinnis2007a,Hu2010,Lin2014,Siria2017}. The kinetic energy of these flows is either lost to friction with the pore walls or dumped with the concentrated feed solution, and harvesting it could mitigate the energetic cost of filtration. There also exist various strategies for converting industrial waste heat into liquid flow~\cite{Barragan2017,Pascual2023}. Despite the ubiquity of waste flows, they have so far been converted to useful work only with limited efficiency. 

Existing strategies are based on a wide range of astute physical principles, such as droplet impacts and triboelectricity~\cite{Xie2014, Nie2019,Xu2020, Riaud2021}, water evaporation \cite{Yang2017,Ding2017,Xue2017,Dao2020}, or electrolyte flows along conductors~\cite{Dhiman2011,Yin2014,Rabinowitz2020,Chen2023,Xiong2023}, to name a few.
Although diverse, these methodologies are rooted in the streaming of dissolved ions under the liquid flow, which -- through the image charge effect -- results in an electronic current in the contiguous conducting wall: an effect that has been termed "ionic Coulomb drag". 
These routes led to the design of small-scale energy generators, with various applications. However, their power-density remains overall limited, as well as difficult to scale up. In particular, ion polarization effects at membranes, and the charge imbalance associated with ion separation, are strong limitations to any ion-based energy conversion process \cite{Wang2022}. 

\newpage
\noindent{\bf \large Ion-free hydrodynamic Coulomb drag} \\
However, the use of ions as intermediate charge carriers to induce an electronic current is not necessarily a prerequisite. Indeed, we have shown recently ~\cite{Coquinot2023} that it is possible to \emph{directly} convert the kinetic energy of a flowing liquid  into an electronic current within the solid wall -- a phenomenon dubbed hydrodynamic Coulomb drag~\cite{Coquinot2023} or hydro-electronic drag. In the following, we adopt the latter denomination, to emphasize the distinction with the ionic Coulomb drag described above. At the root of hydro-electronic drag is the solid-liquid quantum friction phenomenon~\cite{Kavokine2022,Bui2023,Yu2023,Coquinot2023b,Kavokine2022b}, where fluctuating interactions between the liquid and the solid result in momentum transfer from the liquid to the solid's electrons. The rate of this momentum transfer is quantified by the quantum or hydro-electronic friction coefficient, expressed as \cite{Kavokine2022}
\begin{equation}
\lambda_{\rm he} = \frac{1}{8 \pi^2} \int_0^{+\infty}q \mathrm{d} q \, (\hbar q)\int_0^{+\infty} \frac{\mathrm{d}( \hbar \omega)}{k_{\rm B} T} \frac{q}{ \mathrm{sinh}^2\left(\frac{\hbar \omega}{2k_{\rm B} T}\right)}\frac{ \mathrm{Im}[g_{\rm e}(q,\omega)] \, \mathrm{Im}[g_{\rm h}(q,\omega)] }{|1-g_{\rm e}(q,\omega)\,g_{\rm h}(q,\omega)|^2},
\label{qfriction}
\end{equation}
where the functions $g_{\tn{h/e}}$ are the surface response functions of the fluid and electrons in the solid, respectively, describing their charge density fluctuations~\cite{Kavokine2022}. Quantum friction accounts, in particular, for the anomalous water permeability of carbon nanotubes, in terms of the anomalously large quantum friction on multilayer carbon surfaces as compared to isolated graphene layers \cite{Secchi2016,Kavokine2022}. 

Overall, a liquid flow can thus transfer momentum to the solid's electrons, resulting in an electric current proportional to the flow velocity. This process requires neither dissolved ions in the liquid, nor a surface charge on the solid wall. It is thus a mechanism very different from ionic Coulomb drag, and rather analogous to the "condensed-matter" Coulomb drag between two solid-state conductors~\cite{Narozhny2016}.

\rev{We expect hydro-electronic drag to be one of the mechanisms at play in the various experiments that have revealed, in one way or another, an interaction between liquid or ion flow and solid-state electronic current. Unfortunately, many of these experiments have only been carried out with ionic solutions~\cite{Ghosh2003,Dhiman2011,Rabinowitz2020,Chen2023}, making it difficult to disentangle the contributions of ionic and hydro-electronic drag. A few experiments so far have shown electric current generation by ion-free liquids at the mesoscopic scale~\cite{Lee2013,HoLee2013}. While qualitatively in line with hydro-electronic drag theory, the results are difficult to assess quantitatively, as the experiments did not precisely control the interfacial flow velocity; moreover, they show non-linear flow velocity dependence, which points to mesoscopic effects that cannot be understood at the scale of the interface. Recently, however, our own experiments at the micron scale~\cite{Lizee2023} could be quantitatively explained by a phonon-mediated version of hydro-electronic drag}. 

\rev{For a well-chosen solid liquid system, the intrinsic hydro-electronic drag may exceed the contribution of dissolved ions}. Indeed, the number of electrons set in motion through ionic Coulomb drag cannot exceed the number of ionic charges at the solid-liquid interface. Conversely, the liquid's charge fluctuations -- dubbed "hydrons"~\cite{Yu2023,King2023} --, when biased by the hydrodynamic flow, may set in motion all of the conduction electrons within a certain skin depth from the surface (see SI Sec. III). \rev{According to Eq. (1), this requires a frequency matching between the hydron modes and the solid's electronic excitations.}
 To emphasize the crucial role played by hydrons, we will refer to the electrical energy produced through hydro-electronic drag as \emph{hydronic energy}. 

Motivated by these promising qualitative features, we undertake in this Article to quantitatively assess the efficiency of hydronic energy conversion. To this end, we develop a non-equilibrium thermodynamic formalism for the solid-liquid interface, and derive a dimensionless hydronic figure of merit that controls the conversion efficiency. We find that, in practical cases, this figure of merit can significantly exceed unity, highlighting the potential of hydro-electronic drag as a new physical principle for energy harvesting from waste flows. 

\begin{figure*}
\centering	
\includegraphics[width=\textwidth]{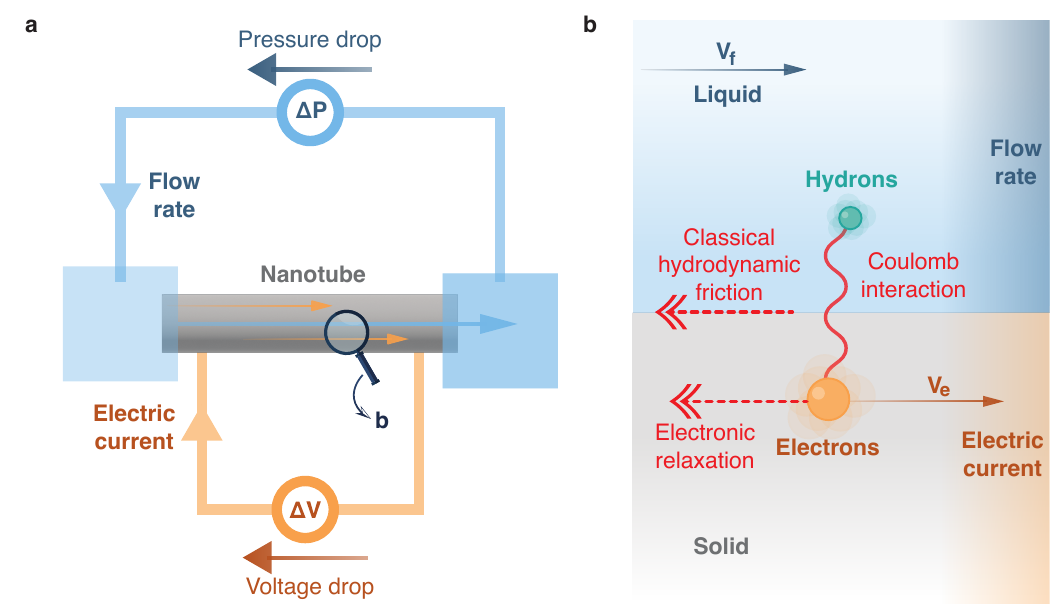}
\caption{\tb{Coupled fluid and electron transport at the solid-liquid interface.}
\tb{(a)} Sketch of a hydronic generator: a liquid flows through a nanotube, connected to an external electric circuit. Thanks to the hydro-electronic drag effect, an electric current is induced in the circuit in response to the liquid flow. \tb{(b)} Schematic of the momentum transfer and relaxation processes at the solid-liquid interface, at the basis of hydro-electronic drag effect. Quantum (or hydro-electronic) friction transfers momentum directly from the liquid to the solid's electrons. Momentum is relaxed at the interface through classical hydrodynamic friction, and inside the solid through electron scattering.}
\end{figure*}

\vskip0.5cm
\noindent{\bf \large Modelling a hydronic generator} \\
 We consider the elementary building-block of an energy conversion device based on hydro-electronic drag -- which we dub ``hydronic generator''. It consists in a nanoscale tube of length $L$, radius $a$ and thickness $\delta \ll a$, connected mechanically to two fluid reservoirs, and electrically to an external circuit that allows for electron circulation through the tube wall (Fig. 1). A pressure drop $\Delta P$ may be applied between the two reservoirs, and a voltage drop $\Delta V$ may be applied between the two solid-state electrical contacts. The fluid, with viscosity $\eta$, is assumed incompressible and the Reynolds number is much smaller than 1, so that the flow rate through the tube in the absence of entrance effects is given by the Poiseuille law~\cite{Kavokine2021}:
\beq 
Q=\frac{\pi a^4}{8\eta}\frac{\Delta P}{L}+\pi a^2 v_{\rm h}.
\eeq
Here we introduced a slip velocity $v_{\rm h}$ for the fluid at the interface, which remains to be determined from the boundary conditions. 

The electric current flowing through the solid wall can be rigorously determined from the electron's non-equilibrium Green's function renormalized by the water-electron Coulomb interactions, as was done in ref.~\cite{Coquinot2023} using Keldysh perturbation theory. Here, we use instead a simplified Drude model, which is quantitatively accurate for systems with a simple band structure~\cite{Coquinot2023}, and has the advantage of being readily integrated into a thermodynamic formalism. The electric current is then $I = - 2 \pi a \delta \times n_{\rm e} e v_e$, where $n_{\rm e}$ where is the electron density and $v_{\rm e}$ is the electron drift velocity, assumed uniform throughout the solid. In the Drude framework, we assume the electrons to have a parabolic dispersion with effective mass $m_{\rm e}$ and a momentum-independent relaxation time $\tau_{\rm e}^0$. A force balance on the electrons then yields 
\beq
I = \frac{n_{\rm e} e^2 \tau_{\rm e}^0}{m_{\rm e}} \Delta V - \frac{1}{L} \frac{e \tau_{\rm e}^0}{m_{\rm e}} F_{\rm he},
\eeq
where $F_{\rm he}$ is the force exerted by the fluid on the electrons, which remains to be specified. In the following we will use the notation $\lambda_{\rm e}^0 \equiv  n_{\rm e} m_{\rm e}\delta /\tau_{\rm e}^0$, a measure of electron relaxation that has the dimension of a friction coefficient. We emphasize again that the assumptions of the Drude model could be relaxed at this point~\cite{Coquinot2023}, however at the expense of simplicity.

The flowing liquid transfers momentum to the channel wall through hydrodynamic friction. This momentum is redistributed between the wall's various degrees of freedom (phonons, electrons) and is eventually relaxed to the environment. The total momentum flux (or force) $F_{\rm hs}$ from the liquid to the solid may be phenomenologically separated into two parts: a part that reaches the electrons (and is then dissipated by the electronic relaxation mechanisms) and a part that doesn't. The latter corresponds to the classical (roughness-induced) friction~\cite{Kavokine2022}. The former comprises hydro-electronic friction (Eq.~\eqref{qfriction}) and possibly a part due to phonons~\cite{Kral2001,Lizee2023,Coquinot2023,Lizee2024}. We will not consider the phonon contribution at this stage, and will thus provide lower bounds for the hydro-electronic coupling effects, in the absence of phononic enhancement. Quantitatively, 
\beq
F_{\rm hs}/(2\pi a L) = \lambda_{\rm h}^0 v_{\rm h}+ \lambda_{\rm he} (v_{\rm h} - v_{\rm e}) , 
\eeq
where $\lambda_{\rm h}^0$ is the classical friction coefficient and $\lambda_{\rm he}$ the hydro-electronic (quantum) friction coefficient, introduced in full generality in Eq. (\ref{qfriction}).
We then identify the hydro-electronic force as $F_{\rm he} = 2\pi a L \lambda_{\rm he} (v_{\rm h} - v_{\rm e})$. Finally, by enforcing global force balance on the liquid, $F_{\rm hs} = \pi a^2 \Delta P$, we obtain a closed set of equations, that can be solved to obtain the "fluxes" $Q$ and $I$ as a function of the "forces" $\Delta P$ and $\Delta V$.

\begin{figure*}
\centering	
\includegraphics[width=\textwidth]{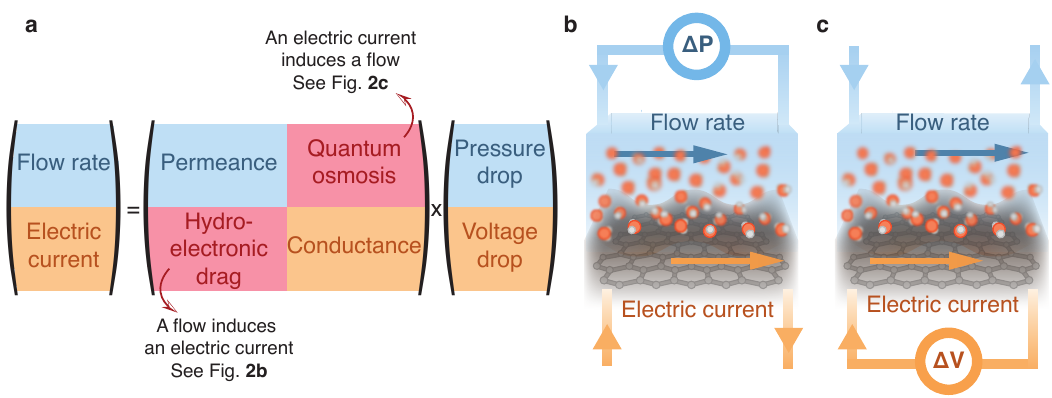}
\caption{\tb{Transport matrix and hydro-electronic cross-couplings.}
\tb{(a)} Sketch of the transport matrix of the system including a coupling (in red) between the flow sector (in blue) and the electronic sector (in orange). The two cross-terms correspond respectively to the hydro-electronic drag and quantum osmosis effects, and the corresponding transport coefficients are equal by Onsager's symmetry.
\tb{(b)} Sketch of the hydro-electronic drag effect: a pressure drop generates a flow which induces an electric current through the hydro-electronic friction. 
\tb{(c)} Sketch of the quantum osmosis effect: a voltage drop generates an electric current which induces a flow through the hydro-electronic friction.}
\end{figure*}

\vskip0.5cm
\noindent{\bf \large Hydro-electronic transport matrix} \\
The solution adopts a matrix structure (see SI Sec. I):
\beq   \begin{pmatrix} Q \\ I \end{pmatrix} = \tb{L}\begin{pmatrix}  \Delta P \\ \Delta V \end{pmatrix} \quad \tn{ with }\quad  \tb{L}= \begin{pmatrix} \mc{L} & \Cef \\ \Cef & G \end{pmatrix}
\label{matrixT}\eeq 
as sketched in Fig. 2a. The diagonal elements of the transport matrix are the permeance $\mc{L}$ and the conductance $G$. We find that the off-diagonal elements $\Cef$ are non-zero and equal, as expected from Onsager symmetry. Note that $\Cef$ is directly proportional to the "electro-fluidic conductivity" introduced in ref.~\cite{Coquinot2023}.
Our model therefore not only predicts the hydro-electronic drag (a liquid flow induces an electric current, as sketched in Fig. 2b), but also its reciprocal effect (an electric current induces a liquid flow, as sketched in Fig. 2c), which has so far never been observed, and which we dub "quantum osmosis". In the following, we will call $\Cef$ the "hydro-electronic mobility". 

To provide convenient expressions of the transport coefficients, we define a dimensionless hydro-electronic coupling constant
\beq
 \Lambef=\frac{\lambef^2}{(\lambf^0+\lambef)(\lambe^0+\lambef)}<1.
\eeq
$\Lambef$ quantifies the competition between liquid-electron momentum transfer ($\lambef$) and the various sources of momentum relaxation: classical hydrodynamic friction ($\lambda_{\rm h}^0$) and electron scattering ($\lambda_{\rm e}^0$).

Accordingly, the expression for the hydro-electronic mobility $\Cef$ defined in Eq. (\ref{matrixT}) -- at the core of the present study -- takes the form
\beq 
\Cef=\frac{\pi a^2\delta}{L} \times  \frac{en_{\rm e}}{\lambef} \times \frac{\Lambef}{1-\Lambef}.
\eeq
It vanishes in the absence of hydro-electronic friction ($\Lambef=0$) and grows to infinity as $\Lambef\rightarrow$  1. 

The formalism also provides expression for the diagonal terms of the transport matrix: the permeance, which determines the flow rate $Q$ under a pressure drop $\Delta P$ (for $\Delta V=0$)
\beq
 \L= \frac{\pi a^4}{8 \eta L} \left( 1 + \frac{4 b}{a (1 - \Lambef)} \right),
 \eeq
where $b \equiv \eta / (\lambda_{\rm ef} + \lambf^0)$ is the hydrodynamic slip length; and the conductance, which determines the electric current $I$ under a voltage drop $\Delta V$ (for $\Delta P=0$)  
\beq 
 G=\frac{1}{1-\Lambef}\times\frac{2\pi a \delta}{L}\times\frac{e^2 n_{\rm e}\taue}{m_{\rm e}},  
 \eeq
where $\tau_{\rm e}^{-1} = (\lambe^0 + \lambef) / \delta n_{\rm e} m_{\rm e}$ is the total electron scattering rate. These expressions highlight that the hydro-electronic coupling modifies the usual pressure-driven and voltage-driven transport phenomena. In the absence of hydro-electronic coupling ($\Lambda_{\rm he} = 0$) the permeance and the conductance reduce to the Poiseuille formula and the Drude formula, respectively. When $\Lambda_{\rm he} \neq 0$, we find that both transport coefficients are enhanced. Physically, the flow-induced electric current boosts the liquid flow along the wall. Conversely, the current-induced liquid flow makes the electrons move faster inside the wall. 

Overall, we have derived a new transport matrix formalism the solid-liquid interface, which bears analogy with the theoretical descriptions of osmotic effects in solution~\cite{Marbach2019} or thermoelectric effects in the solid state \cite{DiSalvo1999,Shi2020}. We may check explicitly that the transport matrix $\tb{L}$ is definite positive, so that our model satisfies the second law of thermodynamics. We may also invert it in order to determine the voltage induced by the liquid flow in the case where the electric circuit is open, or the hydrostatic pressure induced by the electric current if the channel is closed (see SI Sec. I). 

\rev{At this point, we can provide a first quantitative estimate of the hydro-electronic drag effect. As an example, we take a multiwall carbon nanotube of radius $a = 40~\rm nm$, which was predicted to display significant hydro-electronic friction ($\lambef/\lambf^0 \approx 25$)~\cite{Kavokine2022}. The closed-circuit current and open-circuit voltage across the tube under a pressure drop of 1 bar are displayed in Fig. 3a as a function of load resistance. At vanishing load resistance, the predicted electric current is 40 pA: the effect may thus be measurable at the scale of a single tube. }

Let us now use the transport matrix to determine the efficiency of the hydronic generator. 

\begin{figure*}
\centering	
\includegraphics[width=\textwidth]{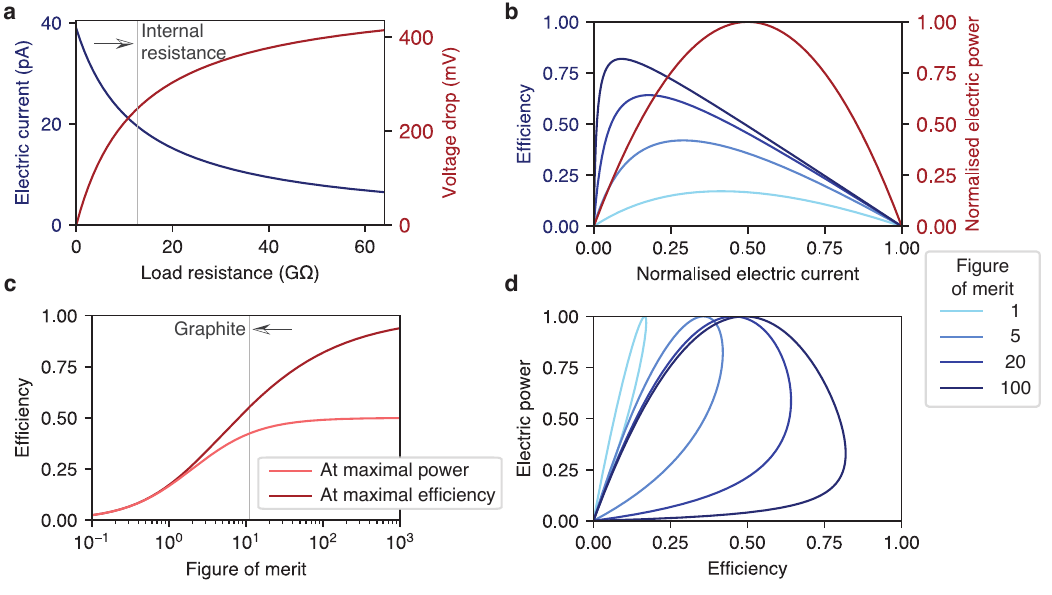}
\caption{\tb{Energetics of the hydronic generator}.
\rev{\tb{(a)} Open-circuit current and closed-circuit voltage produced in the the hydronic generator as a function of the load resistance $R_L$. Here, we consider a single multiwall CNT with radius $a = 40~\rm nm$ and length $L = 10~\rm \mu m$ under a pressure drop of 1 bar. We use $n_{\rm e}=2.3\e{12}$ cm$^{-2}$ for the electron density~\cite{Kavokine2022}, and deduce an internal resistance $G^{-1}\approx 13$ $G\Omega$ for the nanotube, which is indicated by a vertical line.}
\tb{(b)} Efficiency $\gamma$ (in blue) and normalized electric power $\Pe/\Pe^{\rm max}$ (in red) as a function of the normalized electric current $\mc{I}$. The efficiency is displayed for several values of the figure of merit $Z$; the normalized electric power is independent of the figure of merit. 
\tb{(c)} Maximal efficiency $\gamma_{\rm max}$ and efficiency at maximal power ($\mc{I}=1/2$) as a function of the figure of merit $Z$. 
\rev{The efficiency estimate for a graphite-based generator is indicated by a vertical line. }
\tb{(d)} Efficiency-power diagram of the hydronic generator for different figures of merit $Z$. A loop-shaped curve is obtained as the load resistance is swept from zero to infinity. }
\end{figure*}

\vskip0.5cm
\noindent{\bf \large Efficiency and figure of merit} \\
The mechanical power required to impose a flow rate $Q$ through the hydronic generator is $\Pf = Q \Delta P$. The transport matrix yields the corresponding pressure drop $\Delta P$ and the induced electric current $I$ (see SI Sec. IV). The resulting voltage drop is determined by the external load resistance $R_L$ connected to the electric circuit: $\Delta V = - R_L I$. The electric power delivered to the load resistance is then $\Pe = -I \Delta V$, and the hydroelectric energy conversion efficiency is defined as $\gamma = \Pe/\Pf$. 

\rev{The efficiency depends on the magnitude of the load resistance (to be compared with the internal resistance of the system $G^{-1}$), or equivalently on the electric current circulating through the system, as displayed in Fig. 3b (see also Fig. 3a).} At vanishing load resistance, the maximum current is $I\m=\Cef \Delta P$. For a given value of $R_L$, the efficiency can be expressed in terms of two dimensionless numbers, the normalized electric current $\mc{I} = I / I\m$ and the figure of merit $Z$: 
\beq
 \gamma =  \frac{\mc{I}(1-\mc{I})}{\mc{I}+1/Z},
 \eeq
with 
\beq
Z =\frac{\lambef^2}{\lambe^0\lambf^0+\lambef(\lambe^0+\lambf^0)}\frac{1}{1+(\lambf^0+\lambef)a/4\eta}. 
\eeq
This definition of the hydronic figure of merit is inspired by its thermoelectric analogue. Physically, $Z$ compares the liquid-electron energy transfer rate ($\lambda_{\rm he}$) to the rate of energy dissipation in the system, which originates from electron scattering ($\lambe^0$), classical hydrodynamic friction ($\lambf^0$), and viscous effects in the fluid ($\eta$). 

The efficiency vanishes when $Z = 0$ and approaches 1 with increasing $Z$, as displayed in Fig. 3c: here, the efficiency is not limited by the Carnot efficiency as there are no temperature gradients involved. For a given $Z$, there is a value of electric current $\mc{I}$ that achieves the maximum efficiency, given by 
\beq
\gamma\m=\frac{Z}{(1+\sqrt{1+Z})^2}=1-\frac{2}{\sqrt{Z}}+O\left(\frac{1}{Z}\right).
 \eeq
 The delivered electric power is conveniently expressed as 
 \beq
  \Pe=\mc{I}(1-\mc{I})Z \mc{L}^0 (\Delta P)^2,
  \eeq
where $\mc{L}^0 = \pi a^4 (1 + 4b/a)/8\eta L$ is the permeance of the channel in the absence of flow-induced electric current. There is a value of $\mc{I}$ that achieves maximum power, which is different from the one that achieves maximum efficiency. This tradeoff is highlighted by the full power-efficiency diagram displayed in Fig. 3d. 

We may now qualitatively analyze the requirements for a large figure of merit $Z$, and contrast them with the case of thermoelectricity. The thermoelectric figure of merit is essentially controlled by the ratio of electrical and thermal conductivity in a given material. These two quantities tend to vary together, which makes it difficult to achieve a figure of merit exceeding unity. In the hydro-electronic case, however, there is no contradiction between maximizing the electron-liquid interaction and minimizing dissipation (essentially, surface roughness and electron-phonon coupling). It is therefore of interest to make a quantitative assessment of the efficiency achievable in a hydronic generator and of the ensuing potential for waste flow recovery. 

\vskip0.5cm
\noindent{\bf \large Waste flow recovery} \\
We first provide a simplified expression of the figure of merit by analyzing the relative importance of the dissipation mechanisms. Typical hydrodynamic friction coefficients are in the range $\lambf^0 \sim 10^3 - 10^6$ Pa.s/m. The electron relaxation (mostly due to electron-phonon scattering at room temperature) is strongly material-dependent, but is typically in the range $\lambe^0 \sim 10^{-2}-10^3$ Pa.s/m (SI Sec. II). Thus, in most practical cases, dissipation is dominated by viscosity and interfacial friction effects and the figure of merit simplifies to 
\beq
Z \approx \frac{\lambda_{\rm he}}{\lambf^0} \times \frac{1}{1+ a (\lambef + \lambf^0)/4\eta}. 
\eeq
This expression involves two physically meaningful quantities: the ratio of the hydro-electronic and classical friction coefficients and the ratio of channel radius and slip length $ b = \eta / (\lambef + \lambf^0)$. The latter reflects that if the channel is large, most of the mechanical power is lost to viscous dissipation inside the fluid, rather than converted to electric power at the interface. Conversely, a sufficiently narrow channel ($a \ll b$) allows for dissipation-free plug flow. The larger $\lambef$, the narrower the channel needs to be to avoid viscous losses. 

\begin{figure*}
\centering	
\includegraphics[width=\textwidth]{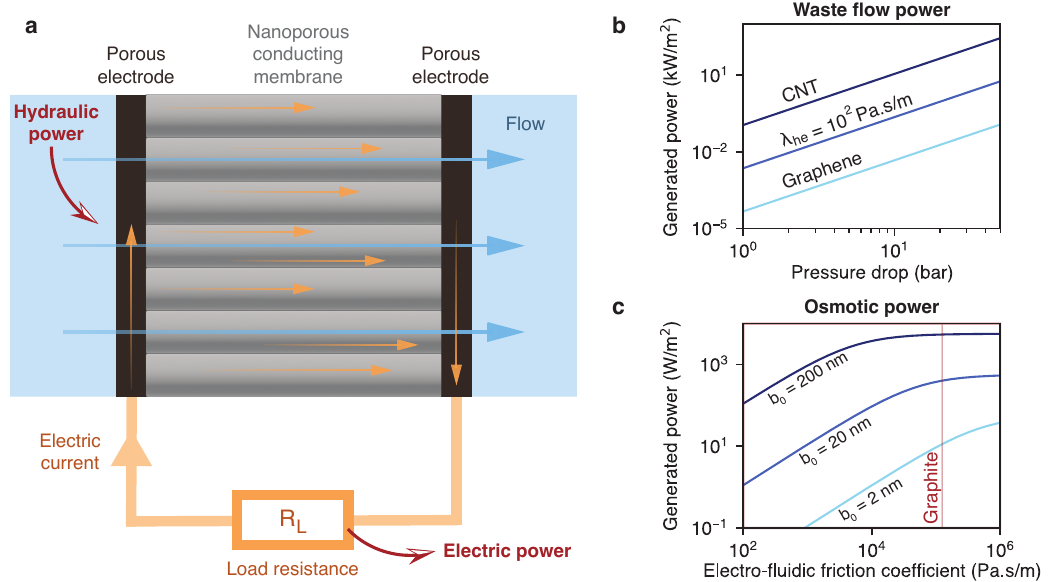}
\caption{\tb{Membrane-scale hydronic generator for waste flow recovery.}
\textbf{(a)} Schematic of a hydronic generator based on a nanoporous conducting membrane placed between two porous electrodes. 
\textbf{(b)} Power recovered with a membrane-scale hydronic generator as a function of waste flow pressure, for different membrane materials. We consider multiwall CNTs with radius $a = 40~\rm nm$, length $L = 10~\rm \mu m$ and 50\% packing fraction. We also consider pores with the same geometric and roughness characteristics, but with lower values of hydro-electronic friction coefficient: $\lambda_{\rm he} = 10^2~\rm Pa.s/m$, and $\lambda_{\rm he} = 1~\rm Pa.s/m$, as predicted for doped graphene~\cite{Kavokine2022}. \textbf{(c)} Power produced by pressure-retarded osmosis through a semi-permeable hydronic generator membrane.  Here, the pores have a radius $a=0.5$ nm and 50\% packing fraction.  The power per unit membrane area is plotted as a function of the hydro-electronic friction coefficient ($\lambef$), for different values of the classical friction coefficient ($b_0 = \eta / \lambda^0_{\rm h}$), and for a membrane thickness $L = 10~\rm \mu m$. The curves flatten at large $\lambef$ due to viscous dissipation in the membrane.}
\end{figure*}

As a case study, let us analyze the performance of a hydronic generator based on a multiwall carbon nanotube (CNT) membrane~\cite{Hinds2004,Holt2006}, as sketched in Fig. 4a. It was shown that, if sufficiently large, these tubes exhibit a specific plasmon mode, which results in significant quantum friction for water: $\lambef/\lambf^0 \approx 25$ for a tube of radius $a = 40~\rm nm$ (where $a/(4b) \approx 1.3$) \cite{Kavokine2022,Secchi2016}. The corresponding figure of merit is indeed significant: $Z \approx 11$, allowing for a maximum efficiency $\gamma\m \approx 0.53$. Thus, a CNT-based hydronic generator could recover more than half of the energy lost to waste flows in a membrane-based filtration process. In reverse osmosis desalination, for example, sea water is pressurized to around 60 bar at the system inlet, and the brine released at the outlet is still at a high pressure, around $10~\rm bar$. Large scale desalination plants implement recovery devices for this brine flow, such as Pelton turbines and hydraulic pressure exchangers~\cite{Schunke2020}. While these devices work with efficiency close to unity in large plants, they are unsuitable for single-household or portable desalination systems. 
\rev{This is a technological gap that could potentially be filled by the hydronic generator. We note, however, that progress towards applications will require specific membrane development, beyond existing solutions.}
CNT membranes are indeed difficult to scale up, there exist many alternative materials that could exhibit similar properties (graphene oxides, lamellar conducting MXenes~\cite{Gogotsi2019}, etc.) and that should be considered as technological pathways to hydronic energy. As a guideline, materials with a relatively low density of high-effective-mass electrons are promising for achieving large figures of merit (SI Fig. 1). For reference, Fig. 4b shows the recovered power per unit area as a function of waste flow pressure for a CNT membrane, as well as for model membrane materials characterized by their hydro-electronic friction coefficient.

A further promising application of the hydronic generator is blue energy harvesting through pressure retarded osmosis (PRO). In PRO, the osmotic pressure difference between fresh water and sea water -- separated by a semi-permeable membrane -- is used to drive a water flow, which is fed into a turbine. Due to the low permeability of the membrane, a flow rate that is sufficient to spin a turbine is achieved only at the power plant scale. \rev{Replacing the turbine with a hydronic generator would make the technology viable at smaller scales: with the CNT-based model system, we estimate an achievable power density of $15~\rm W/m^2$ (see SI Sec. IV), which is larger than what has so far been achieved with a PRO membrane and a turbine ($< 1~\rm W/m^2$) \cite{Logan2012, Siria2017}.} An even higher power density could be achieved with a hypothetical semi-permeable and electrically-conducting membrane, that would simultaneously express the osmotic pressure and harvest the energy from the resulting flow. Indeed, in PRO, most of the energy stored in the salinity gradient is lost to hydrodynamic friction with the membrane. A hydronic generator would instead convert this energy into electricity. \rev{With a semi-permeable membrane that exhibits the same friction characteristics as graphite, a power density exceeding $1~\rm kW/m^2$ could be achieved (Fig. 4c). But even with moderate} hydro-electronic friction ($10^2 - 10^3~\rm Pa.s/m$), a power density exceeding the industrial relevance threshold of $5~\rm W/m^2$ is within reach~\cite{Siria2017}. We note that, in single nanopore systems, while ion-based osmotic power densities up to $1~\rm MW/m^2$ have been demonstrated~\cite{Siria2013, Feng2016}, this figure plummets upon scale-up due to concentration polarization effects and limitations of the electrodes~\cite{Wang2022}. Here, the energy conversion mechanism is ion free, hence it is expected to face fewer scale-up limitations. \rev{Therefore, developing selective membranes with low roughness and high conductivity may a new avenue toward blue energy harvesting.}

\vskip0.5cm
\noindent{\bf \large Perspectives} \\
We have established several unique characteristics of hydro-electronic drag, that set it apart from other hydroelectric energy conversion principles. At the macroscopic scale, turbines are robust and efficient converters, based on inertial effects. As inertia vanishes at smaller scales, "chemical" energy conversion takes over: the energy stored in the fluid is first converted to an ionic current, which is then transformed into an electronic current thanks to an electrochemical reaction. Hydro-electronic drag uses neither inertia nor chemistry, and thus plays the role of a seemingly impossible "nanoscale turbine". 

We have developed a complete thermodynamic formalism for the hydro-electronic cross-couplings at the basis of this nanoscale turbine. A strong analogy can be established with thermoelectric cross-couplings: the induction of an electric current by a liquid flow is reminiscent of the Seebeck effect, while the induction of a liquid flow by an electric current (quantum osmosis) is analogous to the Peltier effect. But a key difference with thermoelectricity is that the relevant figure of merit is controlled by largely independent parameters, and can therefore significantly exceed unity -- a peculiarity rooted in the interfacial nature of hydro-electronic effects. 
Thanks to this peculiarity, hydro-electronic drag is a promising principle for harvesting energy from low Reynolds number waste flows, particularly those that arise in small-scale desalination and pressure-retarded osmosis. Technologies based on hydro-electronic drag will place constraints on membrane materials that are very different from those in electrochemical technologies, as the materials' electronic excitations will come into play. Our results suggest that there are practical consequences for the water-energy nexus of the fundamental fact that, at the nanoscale, classical fluid dynamics meet the quantum dynamics of electronic matter. 

\vskip0.5cm
\noindent{\bf \large Data, Materials and Software availability} \\
All data are included in the manuscript and/or supplementary information.

\vskip0.5cm
\noindent{\bf \large Acknowledgements} \\
The authors thank Mathieu Liz\'ee and Damien Toquer for fruitful discussions. The authors acknowledge support from ERC project {\it n-AQUA}, grant agreement $101071937$. B.C. acknowledges support from the CFM Foundation. The Flatiron Institute is a division of the Simons Foundation.  

\vskip0.5cm
\noindent{\bf \large Author contributions} \\
B.C. developed the theoretical model. All authors were involved in designing the project and writing the paper. 

\vskip0.5cm
\noindent{\bf \large Competing interests} \\
The authors declare no competing interests. 

\bibliography{bibfile}

\end{document}


\title{\textsc{Supplementary information}\\Hydroelectric energy conversion of waste flows through hydro-electronic drag
}

\author{Baptiste Coquinot${}^{1}$}
\author{Lyd\'eric Bocquet${}^1$}\email{lyderic.bocquet@ens.fr}
\author{Nikita Kavokine${}^{2,3}$}\email{nkavokine@flatironinstitute.org}

\affiliation{${}^1$\!\!\! Laboratoire de Physique de l'\'Ecole Normale Sup\'erieure, ENS, Universit\'e PSL, CNRS, Sorbonne Universit\'e, Universit\'e Paris Cit\'e, 24 rue Lhomond, 75005 Paris, France\\
${}^2$\!\!\! Department of Molecular Spectroscopy, Max Planck Institute for Polymer Research, Ackermannweg 10, 55128 Mainz, Germany \\
${}^3$ \!\!\!  Center for Computational Quantum Physics, Flatiron Institute, 162 5$^{th}$ Avenue, New York, NY 10010, USA}

\date{\today}

\maketitle 

\section{Transport matrix}

\subsection{Computation of the transport matrix} 

 The equation of motion for the liquid is:
 \beq \pi a^2 \times \Delta P=2\pi aL\times[\lambf^0\vf+\lambef(\vf-\ve)].\eeq
 We denote $\lambf=\lambf^0+\lambef$ the total equilibrium friction coefficient. 
 For the electrons, the equation of motion is
  \beq 2\pi ad \times n_{\rm e}e\Delta V=2\pi aL\times [\lambe^0\ve+\lambef(\ve-\vf)].\eeq
where  $\lambe^0=d n_{\rm e}m_{\rm e}/\taue^0$ is the friction coefficient by surface area experienced by the electrons due to the electronic relaxation. We denote $\lambe=\lambe^0+\lambef$ the total equilibrium friction coefficient experienced by the electrons. It corresponds to a total electronic scattering time $\taue$  defined by $\lambe=d n_{\rm e}m_{\rm e}/\taue$.

These equations of motion have a common matrix structure given by
\beq  \begin{pmatrix} \lambf & -\lambef \\ -\lambef & \lambe \end{pmatrix} \begin{pmatrix} \vf \\ \ve \end{pmatrix} = \begin{pmatrix}  a\Delta P/2L \\ edn_{\rm e}\Delta V/L\end{pmatrix}
\label{momentum_balance1}\eeq 
We need to invert this matrix to obtain the fluxes. The determinant of the matrix is
\beq \tn{det}\begin{pmatrix} \lambf & -\lambef \\ -\lambef & \lambe \end{pmatrix}=\lambf\lambe-\lambef^2=\lambf\lambe(1-\Lambef)>0\eeq
where 
\beq \Lambef=\frac{\lambef^2}{\lambf\lambe}<1\eeq
is a dimensionless coefficient counting the efficiency of the solid-liquid coupling. 
Solving this linear system we obtain:
\beq   \begin{pmatrix} \vf \\ \ve \end{pmatrix} =  \frac{1}{1-\Lambef}\begin{pmatrix} \lambf\1 & \lambef/ (\lambf\lambe) \\ \lambef/ (\lambf\lambe) & \lambe\1 \end{pmatrix}\begin{pmatrix} a\Delta P/2L \\ edn_{\rm e}\Delta V/L \end{pmatrix}
\label{transport_eq_loc}\eeq

The flow rate writes 
\beq Q=\frac{\pi a^4}{8\eta}\frac{\Delta P}{L}+\pi a^2 \vf \eeq
Thus, we deduce the permeance
\beq \L=\frac{\pi a^4}{8\eta L}+\frac{1}{1-\Lambef}\frac{\pi a^3/L}{2(\lambf^0+\lambef)}=\frac{\pi a^4}{8\eta L}\left(1+\frac{1}{1-\Lambef}\frac{4b}{a}\right)\label{SI_L}\eeq
where  $b=\eta/(\lambf^0+\lambef)$ is the hydrodynamic slip length. We observe an effective slip length $b_{\rm eff}=b/(1-\Lambef)$ in the absence of voltage drop, that is in a short circuit scenario. The slippage is larger because the electrons are in movement, leading to a cancelling of the hydro-electronic friction for the liquid. However, when the circuit is open, that is $I$ is set to be zero, the slip length returns to $b$ as no movement of electrons is allowed.

The electric current writes
  \beq I=2\pi a d\times en_{\rm e}\ve\eeq
  Thus, the conductance is 
    \beq G=\frac{1}{1-\Lambef}\frac{2\pi a(e dn_{\rm e})^2/L}{\lambe^0+\lambef}=\frac{1}{1-\Lambef}\times\frac{2\pi a}{L} \frac{e^2 dn_{\rm e}\taue}{m_{\rm e}} \eeq
    
    Finally, the cross-term is 
    \beq \Cef=\frac{\pi a^2}{L} edn_{\rm e} \frac{1}{1-\Lambef}\frac{\lambef}{(\lambf^0+\lambef)(\lambe^0+\lambef)}.\eeq
    Putting everything together, we have obtained:
    \beq   \begin{pmatrix} Q \\ I \end{pmatrix} = \tb{L}\begin{pmatrix}  \Delta P \\ \Delta V \end{pmatrix} \quad \tn{ with }\quad  \tb{L}= \begin{pmatrix} \mc{L} & \Cef \\ \Cef & G \end{pmatrix}.\eeq  

 For later applications, let us provide the identities
\beq  \Cef=\frac{a}{2edn_{\rm e}}\frac{\lambef}{\lambf^0+\lambef}G=\frac{2edn_{\rm e}}{a}\frac{\lambef}{\lambe^0+\lambef}\frac{\L}{1+a/4b_{\rm eff}}
\label{SI_Cef}\eeq
and 
\beq \Cef^2=\frac{\Lambef}{1+a/4b_{\rm eff}}\L G. \eeq
In particular, the determinant of the transport matrix is 
\beq \det(\tb{L})=\L G\left(1-\frac{\Lambef}{1+a/4b_{\rm eff}}\right)>0\eeq
which is indeed positive. Therefore, the matrix $\tb{L}$ is positive and symmetric.


\subsection{Computation of the drag resistance} 

We now consider the case where we generate a flow rate $Q$ through a pressure drop $\Delta P$, which induces a voltage drop $\Delta V$ when the electric circuit is open, \ie $I=0$. 
Thus, the transport equations writes 
    \beq   \begin{pmatrix} Q \\ 0 \end{pmatrix} =  \begin{pmatrix} \mc{L} & \Cef \\ \Cef & G \end{pmatrix}\begin{pmatrix}  \Delta P \\ \Delta V \end{pmatrix}.\eeq  
    The inverse of the transport matrix writes
     \beq   \tb{L}\1=\frac{1}{1-\frac{\Lambef}{1+a/4b_{\rm eff}}} \begin{pmatrix} \mc{L}\1 & -\Cef\1\frac{\Lambef}{1+a/4b_{\rm eff}} \\ -\Cef\1\frac{\Lambef}{1+a/4b_{\rm eff}} & G\1 \end{pmatrix}.\eeq 
     As a consequence, the drag resistance defined by $\Delta V=-R_{\rm D}Q$ is
     \beq R_{\rm D}= \frac{L\Lambef}{1+a/4b_{\rm eff}-\Lambef}\Cef\1\eeq
     Using Eq. \eqref{SI_Cef}, we get 
          \beq R_{\rm D}= \frac{L}{\pi a^2 edn_{\rm e}}\frac{1-\Lambef}{1+a/4b_{\rm eff}-\Lambef}\lambef\eeq
          Finally, using that $b=\eta/(\lambf^0+\lambef)(1-\Lambef)$, we deduce
              \beq R_{\rm D}= \frac{L}{\pi a^2 edn_{\rm e}}\frac{\lambef}{1+(\lambf^0+\lambef)a/4\eta}.\eeq
              
              Lastly, let us notice that the computation is identical when  inverting the electric current with the flow rate by symmetry of the transport matrix. Thus, for a scenario where the the liquid cannot flow, an electric current induces a pressure drop $\Delta P= -R_{\rm D}I$. 
              

\section{Materials}


\subsection{Jellium model} 

\begin{figure*}
\centering	
\includegraphics{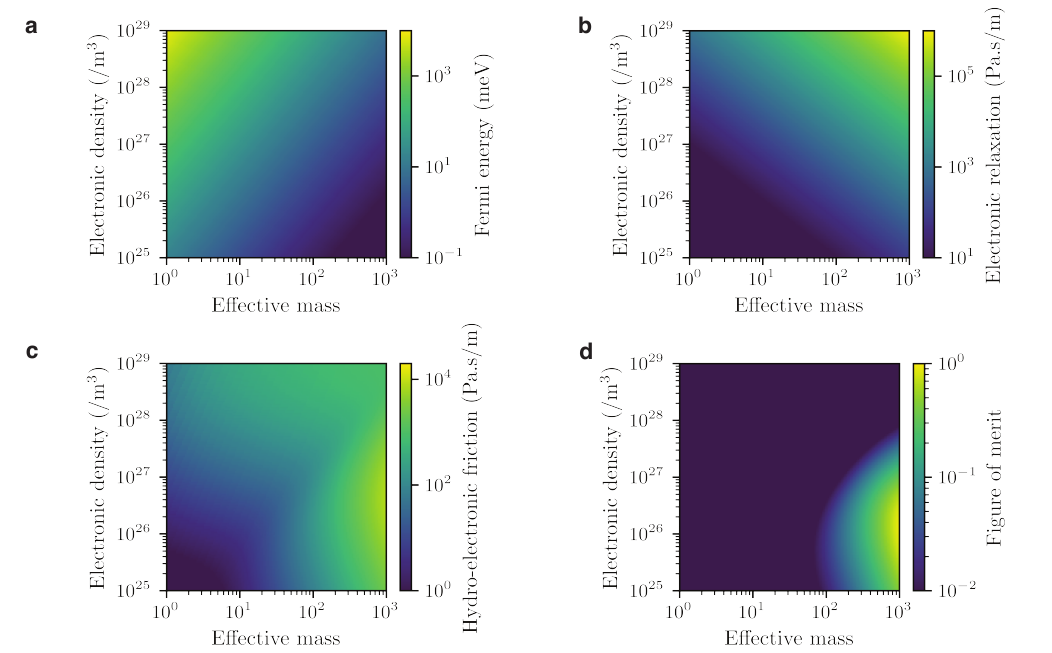}
\caption{\tb{Phase diagram of the Jellium model efficiency for the hydronic generator.} The parameters are the effective mass $m^*$ and the electronic density $n_{\rm e}$. The jellium model is assumed to be at zero temperature for simplicity. 
\textbf{(a)} Effective Fermi energy of the model. 
\textbf{(b)} Electronic relaxation coefficient $\lambe^0$. 
\textbf{(c)} Hydro-electronic friction coefficient $\lambef$.
\textbf{(d)} Associated figure of merit $Z$. 
}
\end{figure*}

A jellium model is made of free electrons with an effective mass $m=m^*m_{\rm e}$ where $m_{\rm e}$ is the electron mass with a positive background \cite{Lang1970}. It is a standard model for the electronic properties of metals and semi-conductors. The energy is $E(k)=\hbar^2k^2/2m$ and the effective mass tunes the band curvature. The electronic density $n_{\rm e}$ fixes the Fermi energy $E_{\rm F}=(\hbar^2/2m)\times (3\pi^2n_{\rm e})^{2/3}$. To keep computations tractable, we consider the jellium at zero temperature, keeping in mind that the results will be qualitative rather than quantitative when the Fermi energy is comparable to temperature~\cite{Kavokine2022}. Globally, we observe that the hydro-electronic friction coefficient becomes larger with larger effective mass and lower electronic density.

The electronic relaxation is $\lambe^0=dn_{\rm e} m/\tau_e^0$ where $\tau_e^0$ is the electronic scattering time. As an order of magnitude, we will take $\tau_e^0 \sim 0.1~\rm ps$, which is typical for graphene~\cite{Coquinot2023}. Using a thickness $d=$ 1 nm, we obtain 
\beq \lambe^0\sim 10\times m^*\times  n_{\rm e} \quad \tn{[}n_{\rm e}\tn{ in /nm}^3\tn{]}\eeq 
Thus, the effective mass and the electronic density should not be too high in order to avoid the electronic relaxation to become important. 

The hydro-electronic friction coefficient, electronic relaxation coefficient and figure of merit (see Sec. IV) for a jellium model are given in Fig. 1. The best candidates are metallic systems with moderate electronic density and flat bands (high effective mass). For example, a jellium model with an electronic density $n_{\rm e}\sim 10^{26}$/m$^3$ and an effective mass $m^*=100$ would have an hydro-electronic friction coefficient $\lambef\approx 2\e{2}$ Pa.s/m and an electronic relaxation $\lambe^0\approx 1\e{2}$ Pa.s/m $\ll \lambef$. Assuming a smooth solid with slip length $b_0\approx 200$ nm like on graphene and no viscous dissipation ($a\ll b$), this corresponds to a figure of merit of $Z\approx 4\e{-2}$. If the effective mass is boosted to $m^*=1000$, the hydro-electronic friction coefficient booms to $\lambef\approx 6\e{3}$ Pa.s/m with an electronic relaxation $\lambe^0\approx 1\e{3}$ Pa.s/m $\ll \lambef$. This leads to a figure of merit of $Z\approx 1$.


\subsection{Numerical values for different physical materials}

\paragraph{Graphene.} The friction on atomically smooth graphene is weak. While there are large discrepancies in the literature for the slip length on graphene, we can use the conservative value of $b=200$ nm. The hydro-electronic friction on graphene is very low. According to \cite{Kavokine2022}, for a carrier density of $n_{\rm e}^{2d}\approx 10^{14}$/cm$^2$, the hydro-electronic friction coefficient becomes $\lambef\approx 1$ Pa.s/m. Thus, the friction of graphene is dominated by the roughness-induced friction: $\lambf^0\approx 5\e{3}$ Pa.s/m. The electronic mobility is graphene is high, typically $\mu\approx 2\e{1}$ C.s/kg. Thus, for a carrier density of $n_{\rm e}^{2d}\approx 10^{14}$/cm$^2$ (strongly doped graphene) we deduce $\lambe^0\approx 10^{-2}$ Pa.s/m. The corresponding figure of merit is $Z\approx 4\e{-4}$. 

\paragraph{Graphite.} The classical friction on atomically smooth graphite is the same than on graphene, thus $\lambf^0\approx 5\e{3}$ Pa.s/m. However, the slip length on graphite is measured to be $b=8$ nm. This strong friction was associated to the the hydro-electronic coupling with the plasmon of graphite \cite{Kavokine2022}. We thus deduce $\lambef\approx 1.3\e{5}$ Pa.s/m. At room temperature, the effective surface charge carrier density of graphite is estimated to $n_{\rm e}^{2d}\approx 2.3\e{12}$/cm$^2$. The in-plane conductivity of graphite is measured to be $\sigma\approx 2\e{4}$ C$^2$.s/kg.m$^3$. For a Drude model, $\sigma\approx n_{\rm e}^{2d} e^2\tau_{\rm e}^0/\delta_{\rm il} m_{\rm e}$ where $\delta_{\rm il}\approx 0.335$ nm is the inter-layer distance. Thus, $\lambe^0\approx N_{\rm layer}(en_{\rm e}^{2d})^2/\sigma\delta_{\rm il}\approx 2 N_{\rm layer}$ Pa.s/m, which remains low even for a large number of layers $N_{\rm layer}$. For graphite, the figure of merit is $Z\approx 11$.

\paragraph{Copper.} Assuming one electron per atom, the carrier density of copper is $n_{\rm e}\approx 10^{29}$/m$^3$. The conductivity of clean copper is $\sigma\sim 6\e{7}$ C$^2$.s/kg.m$^3$. Thus, $\lambe^0\approx \delta(en_{\rm e})^2/\sigma\approx 4\e{3}$ Pa.s/m for a thickness of 1 nm. For a jellium model \cite{Lang1970}, the effective mass of electrons is close to unity \cite{Fukuchi1956}. The Fermi energy of a jellium model with the same electronic density is around 7 eV, close to the Fermi energy of copper. The resulting hydro-electronic friction, computed using the method of \cite{Kavokine2022}, is around $6\e{1}$ Pa.s/m. 

\paragraph{Aluminium.} Assuming one electron per atom, the carrier density of aluminium is $n_{\rm e}\approx 10^{29}$/m$^3$. The conductivity of clean aluminium is $\sigma\sim 4\e{7}$ C$^2$.s/kg.m$^3$. Thus, $\lambe^0\approx \delta(en_{\rm e})^2/\sigma\approx 6\e{3}$ Pa.s/m for a thickness of 1 nm. For a jellium model \cite{Lang1970}, the effective mass of electrons is close to unity \cite{Harrison1960}. The Fermi energy of a jellium model with the same electronic density is around 7 eV, around half the Fermi energy of aluminium. The resulting hydro-electronic friction, computed using the method of \cite{Kavokine2022}, is around $6\e{1}$ Pa.s/m.


\subsection{Numerical values for permeances} 

\paragraph{Remark on units.} The most common unit in the literature for the permeance is:
\beq 1\tn{ L/m}^2\tn{.h.bar}\approx 2.78\e{-12}  \tn{  m/Pa.s}\eeq

\paragraph{Selective membrane for pressure retarded osmosis (PRO).} Commercial membranes selective to salt currently have a permeance of typically \cite{Lim2021}:
\beq \L^*_h\sim 1-3\tn{ L/m}^2\tn{.h.bar}\eeq

\paragraph{Membrane of graphitic nanotubes.} For a membrane of graphitic nanotubes of nanometric radius $a$, length $L\sim 10~\mu$m and pore density close to compact packing $\phi\sim 1/\pi a^2$, the permeance per area is 
\beq \L^*=\frac{\pi a^4\phi}{8\eta L}\left(1+\frac{1}{1-\Lambef}\frac{4b_{\rm graphite}}{a}\right)\approx\frac{ab_{\rm graphene}}{2\eta L}\approx 10^{-8}\times a \tn{  m/Pa.s [}a\tn{ in nm]}\eeq
where we have used $b_{\rm graphene}\approx 200$ nm \cite{Secchi2016}.

\section{Ionic Coulomb drag vs. Hydrodynamic Coulomb drag} 

Ionic Coulomb drag corresponds to an electronic current induced by the ionic streaming current in solution \cite{Marbach2019}. In presence of a fixed surface charge $\Sigma$ a flow of liquid at velocity $\vf$ will drag the counter-ions in the solution leading to an ionic current $\j_{\rm sc}=-\Sigma\vf$. If the solid is a metal or a semi-conductor, each of these charges has an image charge in the solid. The ionic current then reflects into an electronic current of the image charge, which cannot exceed $\j_{\rm i}=\Sigma\vf$. This is the ionic Coulomb drag when the electronic circuit is in short circuit, \ie there is no electric field.

This current can be compared with the current generated by hydrodynamic Coulomb drag, which is based on the fluctuation-induced coupling between the liquid's dielectric fluctuations -- called hydrons -- and the electrons. Here, there is no ion. The electronic current in short circuit is $\j_{\rm h}=en_{\rm e}^{2d}\ve$ where the 2d- electronic density is $n_{\rm e}^{2d}=n_{\rm e}\delta$ where $\delta$ is the skin length over which the electrons are dragged with the liquid. Using Eq. \eqref{transport_eq_loc}, the electronic velocity can be compared to the liquid's velocity as $\ve=\lambef/(\lambe+\lambef)\times \vf$. Thus, the hydrodynamic Coulomb drag current writes 
\beq \j_{\rm h}=en_{\rm e}\delta\frac{\lambef}{\lambe+\lambef}\vf\eeq
As expected, if the hydro-electronic coupling is weak compared with the electron's scattering, the hydrodynamic Coulomb drag is inefficient and the ionic Coulomb drag dominates. However, we have seen that on many materials of interest, the hydro-electronic coupling dominates the electron's scattering ($\lambe\ll\lambef$), in which case the electron's velocity $\ve$ is closed to the liquid's velocity $\vf$.

Thus, for material exhibiting a good coupling with the liquid's modes, both the liquid and the electrons move at the same velocity. Comparing the ionic to hydrodynamic Coulomb drag then reduce to compare the ionic and electronic densities of charge. Thus, the ratio can be written as an electronic Dukhin number $\tn{Du}_{\rm e}$ defined by:
\beq \frac{\j_{\rm i}}{\j_{\rm h}}\approx \frac{\Sigma}{en_{\rm e}\delta}=:\tn{Du}_{\rm e}\eeq
Here $\delta$ represents the thickness of the solid wall or the penetration depth of the hydro-electronic drag, whichever is larger: we will take the conservative estimate $\delta \approx 1~\rm nm$. Then, for a typical metal (1 electron per atom), we find $n_{\rm e} \delta \approx 10^3 ~\rm mol/L$, so that $\tn{Du}_{\rm e}$ is much smaller than 1 for all practical surface charges, and hydrodynamic Coulomb drag is the dominant effect. For a doped semiconductor ($n_{\rm e}\delta\approx 10^{16}-10^{17}~ \rm m^{-2}$), we find that hydrodynamic Coulomb drag dominates if the surface charge is lower than $\Sigma_c\sim 10~\rm mC/m^2$.

%
%
              

\section{Hydro-electronic generator efficiency} 

\subsection{Efficiency under a pressure drop} 

We now generate a flow with a pressure drop and feed the electric current into a load resistance $R_L$. Thus, $\Delta V=-R_L I$. The transport equations now write
 \beq   \begin{pmatrix} Q \\ I \end{pmatrix} =  \begin{pmatrix} \mc{L} & \Cef \\ \Cef & G \end{pmatrix}\begin{pmatrix}  \Delta P \\-R_L I \end{pmatrix}.\eeq  
Thus, the current writes
\beq I=\frac{I\m}{1+GR_L} \quad\tn{with} \quad  I\m=\Cef\Delta P\eeq
ans saturates at small load resistance. Defining $\mc{I}=I/I\m$ the normalised current we can use it as a free parameter instead of $R_L$. Indeed, 
\beq GR_L=\frac{1-\mc{I}}{\mc{I}}.\eeq

The electric power delivered is 
\beq \Pe=R_LI^2=\frac{R_L}{(1+GR_L)^2}I\m^2\eeq
It reaches a maximum for $R_L=1/G$ and the maximal electric power is
\beq \Pe^{\rm max}=\frac{\Cef^2}{G}\frac{(\Delta P)^2}{4}=\frac{\Lambef}{1+a/4b_{\rm eff}}\frac{\L(\Delta P)^2}{4}\eeq
Using the definition of $\L$ in Eq. \eqref{SI_L}, the maximal electric power is 
\beq \Pe^{\rm max}=\frac{\Lambef}{1-\Lambef}\frac{\pi a^3}{2(\lambf^0+\lambef)}\frac{(\Delta P)^2}{4L}\eeq
Returning to the electric power expressed as a function of the parameter $\mc{I}$, it becomes
\beq \frac{\Pe}{\Pe^{\rm max}}=4\frac{GR_L}{(1+GR_L)^2}=4\mc{I}(1-\mc{I}).\eeq
which is the same equation than for an thermo-electric generator. 

The flow rate is 
\beq Q=\L\Delta P-R_L\Cef I =\left(1-\frac{\Lambef}{1+a/4b_{\rm eff}} (1-\mc{I})\right)\L\Delta P \eeq
The mechanical power delivered by the flow is
\beq \Pf=Q\Delta P=\L(\Delta P)^2-\frac{R_L}{1+GR_L}\Cef^2(\Delta P)^2\eeq
 which can be written
    \beq \Pf=\left(1-\frac{GR_L}{1+GR_L}\frac{\Lambef}{1+a/4b_{\rm eff}}\right)\L(\Delta P)^2 .\eeq
 Using the parameter $\mc{I}$, the mechanical power of the fluid writes
        \beq \Pf=\left(1-\frac{\Lambef}{1+a/4b_{\rm eff}}(1-\mc{I})\right)\L(\Delta P)^2 .\eeq
        
        Thus, the efficiency of the hydronic generator is
\beq \gamma=\frac{\Pe}{\Pf}=\Lambef\frac{\mc{I}(1-\mc{I})}{1+a/4b_{\rm eff}-\Lambef(1-\mc{I})}\eeq
Defining the figure of merit by
\beq \ZT=\frac{\Lambef}{1-\Lambef+a/4b_{\rm eff}}=\frac{\Lambef}{1-\Lambef}\frac{1}{1+(\lambe^0+\lambef)a/4\eta},\eeq
the efficiency becomes
\beq \gamma=\frac{\Pe}{\Pf}=\frac{\mc{I}(1-\mc{I})}{\mc{I}+1/Z}.\eeq
This figure of merit is then the only parameter to control the efficiency. To better understand it, we can express it explicitly in term of friction coefficients. 
\beq \ZT=\frac{\lambef^2}{\lambe^0\lambf^0+\lambef(\lambe^0+\lambf^0)}\frac{1}{1+(\lambe^0+\lambef)a/4\eta},\eeq
It ranges from zero when the hydro-electronic coupling vanishes to infinity when the relaxation processes disappear. 

The maximum of the efficiency is reached for $\mc{I}=1/\sqrt{1+Z}$ and achieves
\beq \gamma\m=\frac{Z}{(1+\sqrt{1+Z})^2}.\eeq
When $Z\rightarrow \infty$, the efficiency goes to 1: the conversion is perfect. The electric power at maximal efficiency is
\beq \frac{\Pe}{\Pe^{\rm max}}=4\frac{\sqrt{1+Z}-1}{1+Z}.\eeq

The efficiency at maximal power is also a key information. The electric power is maximum for $ \mc{I}=1/2$ and the associated efficiency is
\beq \gamma_{\rm eff}=\frac{1}{2}\frac{Z}{Z+2}=\frac{1}{2}\frac{\lambef^2}{2\lambe\lambf-\lambef^2}.\eeq
When $Z\rightarrow \infty$, the efficiency saturates to 1/2 but the maximal electric power explodes. Indeed, 
\beq \Pe^{\rm max}=\frac{\Lambef}{1+a/4b_{\rm eff}}\frac{\L(\Delta P)^2}{4}=Z\frac{\L^0(\Delta P)^2}{4}\eeq 
where 
\beq \L^0=\frac{\pi a^4}{8\eta L}\left(1+\frac{4b}{a}\right)\eeq
is the membrane's permeance in the absence of flow-induced electric current.
In the meantime, the flow rate is:
\beq Q=\left(1+\frac{Z}{2}\right)\L^0\Delta P \eeq
Thus, the electric power in terms of the flow rate writes:
\beq \Pe^{\rm max}=\frac{Z}{\left(Z+2\right)^2}\frac{Q^2}{\L^0}\eeq


\subsection{Efficiency for membranes in series under a pressure drop}

Let us now consider the case where the hydronic generator is a membrane of area $\A_{\rm he}$ and pore density $\phi$. We denote $\L^*=\phi\L$, $\Cef^*=\phi\Cef$ and $G^*=\phi G$ the coefficients per membrane area. The generator is used in series with a second membrane of hydrodynamic resistivity $R_h^*=1/\L_h^*$ per membrane area and an area $\A_h$. The pressure drop $\Delta P$ is applied to the complete system. The pressure drop applied to the generator is then 
\beq \Delta P_{\rm eff}=\Delta P- \frac{R_h^*}{\A_h} Q\eeq
The generated electric current is then used in a load resistance $R_L$. 

The transport equations are:
 \beq   \begin{pmatrix} Q \\ I \end{pmatrix} =  \A_{\rm he}\begin{pmatrix} \mc{L}^* & \Cef^* \\ \Cef^* & G^* \end{pmatrix}\begin{pmatrix}  \Delta P-R_h^* Q/\A_h \\-R_L I\end{pmatrix}.\eeq  
Rearranging these equations, we have:
  \beq   \begin{pmatrix} 1+\frac{ \A_{\rm he}}{\A_h}\mc{L}^*R_h^* &  \A_{\rm he}\Cef^* R_L^* \\ \frac{ \A_{\rm he}}{\A_h}\Cef^* R_h^* & 1+ \A_{\rm he}G^*R_L \end{pmatrix} \begin{pmatrix} Q \\ I \end{pmatrix} = \A_{\rm he}\begin{pmatrix}  \L^*\Delta P \\\Cef^*\Delta P\end{pmatrix}.\eeq  
  Inverting the matrix, the fluxes write
  \beqa
  Q &=& \frac{1+\A_{\rm he}\left(1-\frac{\Lambef}{1+a/4b_{\rm eff}}\right)G^*R_L}{1+\A_{\rm he}G^*R_L+\frac{ \A_{\rm he}}{\A_h}\mc{L}^*R_h^*+\frac{ \A_{\rm he}^2}{\A_h}\left(1-\frac{\Lambef}{1+a/4b_{\rm eff}}\right)G^*R_L\mc{L}^*R_h^*}\A_{\rm he}\L^*\Delta P\\
  I &=&  \frac{1}{1+\A_{\rm he}G^*R_L+\frac{ \A_{\rm he}}{\A_h}\mc{L}^*R_h^*+\frac{ \A_{\rm he}^2}{\A_h}\left(1-\frac{\Lambef}{1+a/4b_{\rm eff}}\right)G^*R_L\mc{L}^*R_h^*}\A_{\rm he}\Cef^*\Delta P= \mc{I}I_{\rm max} 
  \eeqa 
  
  The electric power writes
  \beq \Pe(R_L,R_h^*)=R_L I^2= \frac{\A_{\rm he}\frac{\Lambef}{1+a/4b_{\rm eff}}G^*R_L}{\left[1+\A_{\rm he}G^*R_L+\frac{ \A_{\rm he}}{\A_h}\mc{L}^*R_h^*+\frac{ \A_{\rm he}^2}{\A_h}\left(1-\frac{\Lambef}{1+a/4b_{\rm eff}}\right)G^*R_L\mc{L}^*R_h^*\right]^2}\A_{\rm he}\L^*(\Delta P)^2\eeq
  which is maximal for
  \beq \A_{\rm he}G^*R_L=\frac{1+\frac{ \A_{\rm he}}{\A_h}\mc{L}^*R_h^*}{1+\frac{ \A_{\rm he}}{\A_h}\left(1-\frac{\Lambef}{1+a/4b_{\rm eff}}\right)\mc{L}^*R_h^*}\eeq
  Its maximal value is
  \beq \tilde\Pe(R_h^*)= \frac{\frac{\Lambef}{1+a/4b_{\rm eff}}}{\left[1+\frac{ \A_{\rm he}}{\A_h}\mc{L}^*R_h^*\right]\left[1+\frac{ \A_{\rm he}}{\A_h}\left (1-\frac{\Lambef}{1+a/4b_{\rm eff}}\right)\mc{L}^*R_h^*\right]}\frac{\A_{\rm he}\L^*(\Delta P)^2}{4}
  \eeq
  
Dividing that the area of both membranes, the power per area of membrane is
\beq \tilde\Pe^*=\frac{ \tilde\Pe}{\A_{\rm he}+\A_h}=\frac{\frac{\Lambef}{1+a/4b_{\rm eff}}}{\left[1+\frac{ \A_{\rm he}}{\A_h}\mc{L}^*R_h^*\right]\left[1+\frac{ \A_{\rm he}}{\A_h}\left (1-\frac{\Lambef}{1+a/4b_{\rm eff}}\right)\mc{L}^*R_h^*\right]}\frac{\L^*(\Delta P)^2}{4\left(1+\frac{\A_h}{ \A_{\rm he}}\right)}\label{Pmax_Rh}\eeq
The ratio $\A_{\rm he}/\A_h$ is a parameter which can be optimised. The optimal value for this ratio is obtained from the root of a polynomial of third order. 

In practice, we are interested in situations where $\mc{L}^*R_h^*\gg 1$ and $1-\frac{\Lambef}{1+a/4b_{\rm eff}}\ll 1$. We can identify the two following regimes:
\begin{itemize}
\item If $\A_{\rm he}/\A_h\ll 1/\mc{L}^*R_h^*\ll 1$ then $\tilde\Pe^*\propto \A_{\rm he}/\A_h$.
\item If $1/\mc{L}^*R_h^*\ll \A_{\rm he}/\A_h\ll \min[1/\mc{L}^*R_h^*(1-\frac{\Lambef}{1+a/4b_{\rm eff}}),1]$ then $\tilde\Pe^*\propto 1$.
\item  If $1/\mc{L}^*R_h^*\ll \min[1/\mc{L}^*R_h^*(1-\frac{\Lambef}{1+a/4b_{\rm eff}}),1]\ll \A_{\rm he}/\A_h\ll \max[1/\mc{L}^*R_h^*(1-\frac{\Lambef}{1+a/4b_{\rm eff}}),1]$ then $\tilde\Pe^*\propto \A_h/\A_{\rm he}$.
\end{itemize}
Thus, the optimal value for $\A_{\rm he}/\A_h$ is expected to be between $1/\mc{L}^*R_h^*$ and $1/B=1/\max[1/\mc{L}^*R_h^*(1-\frac{\Lambef}{1+a/4b_{\rm eff}}),1]$. In such a case, Eq. \eqref{Pmax_Rh} simplifies to:
\beq \tilde\Pe^*\approx\frac{\frac{\A_{\rm he}}{\A_h}}{\left[1+\frac{ \A_{\rm he}}{\A_h}\mc{L}^*R_h^*\right]\left[1+\frac{ \A_{\rm he}}{\A_h}B\right]}\frac{\L^*(\Delta P)^2}{4}\eeq
Therefore, the maximal power is obtained for:
\beq \max[\mc{L}^*R_h^*(1-\frac{\Lambef}{1+a/4b_{\rm eff}}),1]\ll \frac{\A_h}{\A_{\rm he}}\ll \mc{L}^*R_h^*\eeq
and reaches
\beq\Pe^{*\rm max}\approx\frac{\L^*_h(\Delta P)^2}{4}\label{Pmax_Rh}\eeq
 Here, the takeaway is that the generator's permeance should be much larger than the permeance of the external membrane in order to produce the maximal power. However, if the generator's hydrodynamic resistance becomes too small, it becomes "transparent" and does not produce any power. These constraints can be managed thanks to the special effect of the hydro-electronic coupling which both dominates the friction and make the conversion. Interestingly, the external load resistance should be much larger than the generator's internal electric resistance in order to increase the generated power.    
The hydraulic power writes
  \beq \Pf=Q\Delta P= \frac{1+\A_{\rm he}\left(1-\frac{\Lambef}{1+a/4b_{\rm eff}}\right)G^*R_L}{1+\A_{\rm he}G^*R_L+\frac{ \A_{\rm he}}{\A_h}\mc{L}^*R_h^*+\frac{ \A_{\rm he}^2}{\A_h}\left(1-\frac{\Lambef}{1+a/4b_{\rm eff}}\right)G^*R_L\mc{L}^*R_h^*}\A_{\rm he}\L^*(\Delta P)^2\eeq
  Thus, the efficiency is
  \beq \gamma=  \frac{\A_{\rm he}\frac{\Lambef}{1+a/4b_{\rm eff}}G^*R_L}{\left[1+\A_{\rm he}\left(1-\frac{\Lambef}{1+a/4b_{\rm eff}}\right)G^*R_L\right]\left[1+\A_{\rm he}G^*R_L+\frac{ \A_{\rm he}}{\A_h}\mc{L}^*R_h^*+\frac{ \A_{\rm he}^2}{\A_h}\left(1-\frac{\Lambef}{1+a/4b_{\rm eff}}\right)G^*R_L\mc{L}^*R_h^*\right]}\eeq
  This efficiency could be optimized. However, it is more interesting for applications to work at maximal power. Thus, focusing on the previous ansatz for $R_L$ and  $\A_{\rm he}/\A_h$, the effective efficiency is 
  \beq 
 \gamma_{\rm eff}\approx \frac{1}{2}\eeq
This is an upper boundary condition which is valid for a perfect conversion $\frac{\Lambef}{1+a/4b_{\rm eff}}=1$.




\subsection{Osmotic energy harvesting with a selective membrane} 

We consider the case of osmotic energy harvesting in pressure retarded osmosis (PRO). We use in series a first selective membrane of permeance $\L_h^*$ and a hydronic generator membrane with figure of merit $Z \gg 1$. A good selective membrane would have a permeance around $\L_h^*\approx 2 \tn{ L/m}^3\tn{.h.bar}$. Thanks to the selective membrane, the separation of fresh and sea water induces an osmotic pressure $\Delta \Pi\approx 30$ bar. Thus, after optimizing of the ratio of membrane areas and external load resistance, one can expect to generate an electric power
\beq\Pe^{*\rm max}\approx\frac{\L^*_h(\Delta \Pi)^2}{4}\approx 15 \tn{ W/m}^2\eeq
with an efficiency  $\gamma_{\rm eff}\approx 50\%$.

\bibliography{bibfile,my_library}